\documentclass[12pt,a4paper]{elsart}
\usepackage{latexsym,amssymb,epsfig,rotating}

\newcommand{\ea}{\mbox{\Large $\rightarrow$}}

\def\0{\phantom{0}}

\begin{document}
\pagenumbering{arabic}
\baselineskip25pt

\begin{center}
{\bf \large Grand Equilibrium: vapour-liquid equilibria by a new molecular 
simulation method} \\

\bigskip
\renewcommand{\thefootnote}{\fnsymbol{footnote}}
Jadran Vrabec\footnote[1]
{author for correspondence, tel.: +49-711/685-6107, 
fax: +49-711/685-7657, \\ email: vrabec@itt.uni-stuttgart.de}, Hans Hasse
\renewcommand{\thefootnote}{\arabic{footnote}} \\
Institut f\"ur Technische Thermodynamik und Thermische Verfahrenstechnik, \\
Universit\"at Stuttgart, D-70550 Stuttgart, Germany 
\end{center}

 
%\begin{abstract}

\vskip3cm
Number of pages: 20

Number of tables: 7

Number of figures: 5

Running title: VLE by Grand Equilibrium

\clearpage
{\bf Abstract}

\noindent
A new molecular simulation method for the calculation of vapour-liquid 
equilibria of mixtures is presented. In this method, the independent 
thermodynamic variables are temperature and liquid composition. 
In the first step, one isobaric isothermal simulation for the liquid phase is 
performed, in which the chemical potentials of all components and their 
derivatives with respect to the pressure, i.e., the partial molar volumes, 
are calculated. 
From these results, first order Taylor series expansions for the chemical 
potentials as functions of the pressure $\mu_i(p)$ at constant liquid
composition are determined. That information is needed, as 
the specified pressure in the liquid will generally not be equal to the 
equilibrium pressure, which has to be found in the course of a vapour 
simulation. In the second step, one pseudo grand canonical simulation for the 
vapour phase is performed, where the chemical potentials are set
according to the instantaneous pressure $p^v$ using the previously 
determined function $\mu_i(p^v)$. In this way, results for the vapour 
pressure and vapour composition are achieved which are consistent to the 
given temperature and liquid composition.
The new method is applied to the pure Lennard-Jones fluid, 
a binary, and a ternary mixture of Lennard-Jones spheres and shows very 
good agreement with corresponding data from the literature.

% Text body
\section{Introduction}
Due to the rapid increase of available computing power, molecular
simulation is developing to become a standard tool in chemical engineering and
related areas. It can provide a better understanding of industrial processes
and the underlying physics on the molecular level. Furthermore, molecular
simulation has the capability to predict phase equilibria more reliably than
classical thermodynamic models. The calculation of vapour-liquid equilibria 
by molecular simulation is a longstanding and important task. 
In the last 15 years a variety of methods for this purpose have been 
presented. 

There are, among others, thermodynamic scaling \cite{valleau1,valleau2},
histogram reweighting \cite{mcdonald1,potoff1}, Gibbs-Duhem integration 
\cite{kofke1,kofke2}, NpT plus test particle method \cite{fischer,vrabec95}, 
various extensions of it to other ensembles \cite{boda0,boda1,boda}, and, 
most widely used, the Gibbs Ensemble Monte Carlo method (GEMC)
\cite{pana1,pana2}. A discussion of the different approaches can be found in 
the reviews \cite{pana3,baus}. 

The GEMC specifies the thermodynamic variables temperature, global
composition, and either global volume or pressure. Both phases are simulated 
simultaneously in separate volumes, between which real particles are 
exchanged in order to ensure phase equilibrium. On the one hand, this set 
of specified thermodynamic variables is in many cases not convenient,
i.e., when bubble points or dew points are needed, 
where temperature or pressure and the composition of one phase is specified. 
On the other hand, the simultaneous simulation of both phases has the 
disadvantage that fluctuations occuring in one phase influence the other
phase directly. This can result in unfavourable dynamic behaviour of the
simulation, e.g., close to the 
critical point fluctuations can lead to phase reversal in the two 
volumes. Furthermore, real particle exchange between the two phases is 
a major cause of statistical uncertainties. If the liquid is dense 
and/or consists of strongly interacting particles, the insertion and 
deletion of real particles is awkward and perturbs the liquid
causing high statistical uncertainties.

In this work, the {\sl Grand Equilibrium method} for the calculation of 
vapour-liquid equilibria of multi-compo\-nent mixtures is presented. 
This method circumvents the above mentioned 
problems. The specified thermodynamic variables are temperature and liquid 
composition, two independent simulations for both phases are performed and 
no exchange of real particles in the liquid phase needs to be done.

The Grand Equilibrium method has some common features with the NpT 
plus test particle method for mixtures \cite{vrabec95,vrabec96}. 
In fact, the treatment of the liquid phase is identical. 
The difference lies in the procedure for the vapour phase, where only one 
simulation has to be performed and no prior 
knowledge of the vapour-liquid equilibrium is neccessary. Furthermore, the present method is 
related to the method of Ungerer {\it et al.} \cite{ungerer1,ungerer2}, 
but avoids introducing direct coupling between the phases.
Similar to the work of Esobedo \cite{escobedo1,escobedo2}, it employs 
pseudo ensemble simulations.

The Grand Equilibrium method is tested in the present work on the pure 
Lennard-Jones fluid, a binary, and a ternary Lennard-Jones mixture and the 
results are compared to literature data obtained by other methods.

\clearpage
\section{Method}

The basic idea of the Grand Equilibrium method is to evaluate the chemical 
potentials in the liquid as functions of the pressure at a given temperature 
and liquid composition and to use these functions to set the chemical 
potentials in a pseudo grand canonical ensemble simulation for the vapour 
phase. The phase equilibrium conditions are directly introduced in the vapour 
simulation, which yields vapour pressure, vapour composition and 
other properties on the dew line.

Let us consider a mixture consisting of an arbitrary number of components 
$i$ = A, B, C, ... and a given set of intermolecular potentials between all 
species. Having set the independent thermodynamic variables temperature $T$ and 
liquid composition ${\bf x}$ = $x_A$, $x_B$, $x_C$, ..., one simulation in 
the isobaric isothermal (NpT) ensemble is carried out at an arbitrarily 
chosen pressure $p^l$, preferrably not too far from
the vapour pressure $p_{\sigma}$. This simulation yields the density $\rho^l$ = 
$\rho(T,{\bf x}, p^l)$ and the enthalpy $h^l$ = $h(T,{\bf x}, p^l)$. 
Applying fluctuation formulae in the NpT ensemble allows the derivatives of the
density and enthalpy with respect to the pressure at constant temperature and
constant composition to be obtained from that simulation. 
The derivative of the density is calculated from \cite{hill}
\begin{equation}
\left(\frac{\partial \rho}{\partial p}\right)_{T,{\bf x}} ~=~
\rho\beta_T ~=~ \frac{1}{kT} \cdot \frac{N}{<V>^2} \cdot \left(<V^2> - <V>^2\right),  
\label{drhodp}
\end{equation}
where $V$ denotes the instantaneous volume, $N$ the total number of particles, 
$k$ the Boltzmann constant, $\beta_T$ the isothermal compressibility,
and the brackets $<$ $>$ indicate averaging in the NpT ensemble.
The derivative of the enthalpy is calculated from \cite{hill}
\begin{equation}
\left(\frac{\partial h}{\partial p}\right)_{T,{\bf x}} ~=~ \frac{1}{N} \cdot
\left[\frac{1}{kT} \cdot \Big( <H><V> - <HV> \Big) + <V> \right],
\label{dhdp}
\end{equation}
where $H$ denotes the instantaneous enthalpy. 

For the present method it is essential to additionally calculate  
the chemical potentials $\mu_i$ and the partial molar volumes 
$v_i$ of all components from this liquid phase simulation. 
This can be done, e.g., by Widom's test 
particle insertion \cite{widom6328}, according to which
\begin{equation}
\mu_i ~=~ \mu^{\mbox{\scriptsize id}}_i(T) + kT \cdot \ln x_i -
kT \cdot \ln <V \cdot {\rm exp}(-\psi_i/kT)>/N,
\label{mui}
\end{equation} 
where $\psi_i$ denotes the potential energy of a test particle of species $i$
and $\mu_i^{id}(T)$ the part of the ideal chemical potential
which only depends on the temperature. It is convenient to define the 
residual chemical potential of component $i$ as
\begin{equation}
\tilde\mu_i ~=~ [\mu_i ~-~ \mu^{\mbox{\scriptsize id}}_i(T)]/kT.
\label{mures}
\end{equation}
Widom's test particle insertion also yields 
the partial molar volumes \cite{heyes} 
\begin{equation}
v_i ~=~ \frac{<V^2 \cdot {\rm exp}(-\psi_i/kT)>}{<V \cdot 
{\rm exp}(-\psi_i/kT)>} - <V>.
\label{vi}
\end{equation} 

It is important to note that the Grand Equilibrium method is not 
tied to Widom's test particle insertion, which is known to yield high 
statistical uncertainties in very dense and/or strongly interacting fluids. 
In such cases, other more sophisticated methods to determine the chemical 
potentials, such as gradual insertion in extended ensembles 
\cite{voron1,voron2,nezbeda9139}, can be applied alternatively.

The complete set of thermodynamic quantities that have to be calculated from one
NpT simulation at given $T$, {\bf x}, $p^l$ in the liquid phase is 
$\rho^l$, $h^l$, $\beta_T^l$, 
$(\partial h/\partial p)_{T,{\bf x}}^l$, $\mu_i^l$, and $v_i^l$. 

With these quantities, the desired chemical potentials, as functions of the
pressure, are obtained from first order Taylor series expansions \cite{vrabec95}
\begin{equation}
\tilde\mu_i(p) ~\approx~ \tilde\mu_i^l ~+~ v_i^l/kT \cdot (p - p^l), 
\label{muilp}     
\end{equation} 
with
\begin{equation}
\left(\frac{\partial\mu_i^l}{\partial p}\right)_{T,{\bf x}} ~=~ v_i^l.
\end{equation}
It should be mentioned that in typical applications, the accuracy of equation 
(\ref{muilp}) is not influenced sig\-ni\-fi\-cant\-ly by neglecting 
the higher order terms, but rather by the accuracy of $\mu_i^l$ and $v_i^l$ 
achieved in the simulation. Furthermore, in a liquid 
phase not too close to the critical point the chemical potentials are only 
weak functions of the pressure.
\footnote{If equation (\ref{muilp}) is needed in the near-critical region 
and no good guess for $p^l$ is available, succesive simulations in the 
liquid phase can be necessary.}

Having determined the $\mu_i(p)$ from the Taylor series expansions, 
according to equation (\ref{muilp}) by one NpT simulation in the 
liquid mixture, one vapour simulation in a pseudo grand canonical ensemble 
(pseudo-$\mu$VT) has to be carried out. In a common grand canonical ensemble 
($\mu$VT) simulation \cite{allen,frenkel,sadus}, the temperature, the volume, 
and the chemical potentials of all components are set as fixed values. 
In the present pseudo-$\mu$VT simulation the temperature and the volume are 
set in the common way. But instead of setting fixed values for the chemical 
potentials, they are set as functions of the instantaneous pressure in the
vapour $p^v$  using equation (\ref{muilp}) to find $\mu_i(p^v)$. Applying 
this procedure, it is ensured that equilibrium between the liquid phase
and vapour phase is imposed.

During a common $\mu$VT simulation the chemical potentials are set through 
insertion and deletion of particles by the comparison between the 
resulting potential energy change and the desired fixed chemical potential. 
In the proposed pseudo-$\mu$VT ensemble, in which the  
values of the chemical potential are found from the function $\mu_i(p^v)$,
the acceptance probability $P_{acc}$ for particle insertion of species $i$
writes as
\begin{equation}
\!\!\!\!\!\!\!\!\!\!\!\!P_{acc}(N \ea N+1) = {\rm min} \left(1,~
\frac{V}{N_i+1} \cdot
\exp\{\tilde\mu_i(p^v)-\left[U(N+1)-U(N)\right]/kT\}\right),~~
\end{equation}
where $N_i$ denotes the number of particles of species $i$ and $U(N+1)$ the 
configurational energy of the system with $N+1$ particles.
The acceptance probability for particle deletion of species $i$ writes as
\begin{equation}
\!\!\!P_{acc}(N \ea N-1) = {\rm min} \left(1,~ \frac{N_i}{V} \cdot
\exp\{-\tilde\mu_i(p^v)+\left[U(N)-U(N-1)\right]/kT\}\right).~~
\end{equation}
Starting from an arbitrarily chosen low density state point, the vapour 
simulation is forced by this procedure to change its state to the 
corresponding phase equilibrium state point on the dew line.
The vapour pressure $p_{\sigma}$ and the vapour composition on 
the dew line {\bf y} = $y_A$, $y_B$, $y_C$, ... are then simple ensemble 
averages in the vapour simulation in the pseudo-$\mu$VT ensemble. The
same holds for other properties on the dew line, such as dew density and
dew enthalpy.
%\footnote{It should be noted, that there is a difference in the calculation 
%of the configurational enthalpy for different simulation techniques.
%In the case of molecular dynamics the kinetic energy of the 
%     has to be considered, but not in the case of Monte Carlo.}

Experience shows that the pseudo-$\mu$VT simulation moves rapidly into the 
vicinity of the phase equilibrium state point. This process happens
in the equilibration period. During the production period the 
pressure just fluctuates around its constant average.
These inherent fluctuations in the pressure impose fluctuations on the set 
chemical potentials through equation (\ref{muilp}).
Such an interdependence characterizes pseudo statistical ensembles.

Figure 1 gives an impression of a typical equilibration process obtained 
using the procedure described above for a binary 
Lennard-Jones mixture. After an initial
equilibration of 1 000 Monte Carlo loops in the canonical ensemble (NVT), 
a second equilibration in the pseudo-$\mu$VT ensemble is started. It can be 
observed that the instantaneous values of pressure, density, and vapour 
composition reach the vicinity of equilibrium after only 1~000 Monte Carlo 
loops. Due to the experience with various mixtures including nonspherical and 
polar fluids \cite{stoll2001mix}, 
it can be stated that this process is very robust and independent on the 
initial density and composition in the vapour simulation. 

With the knowledge of the vapour pressure, other thermodynamic properties 
of interest on the bubble line can be calculated easily, e.g., the bubble 
density 
\begin{equation}
\rho' ~\approx~ \rho^l ~+~ \rho^l\beta_T^l \cdot (p_{\sigma} - p^l),
\end{equation} 
the bubble enthalpy
\begin{equation}
h'    ~\approx~ h^l    ~+~ \left(\frac{\partial h^l}{\partial p}\right)_{T,{\bf x}} \cdot (p_{\sigma} - p^l),  
\end{equation} 
and the chemical potentials in phase equilibrium
\begin{equation}
\tilde\mu_{i\sigma} ~\approx~ \tilde\mu_i^l ~+~ v_i^l/kT \cdot (p_{\sigma} - p^l).
\end{equation} 

The present method can be applied for the calculation of dew 
points as well. Therefore the phases simply have to be inverted, i.e. one 
NpT simulation in the vapour phase and one pseudo-$\mu VT$ simulation in 
the liquid phase have to be performed. In this case it faces the 
awkward insertion and deletion of real particles in the liquid phase, 
but still has the advantage of independent simulations runs.

Furthermore, the Grand Equilibrium method can be applied to the more 
challenging liquid-liquid equilibria as well. First results \cite{bodaneu} 
confirm the applicability and demonstrate high accuracy.

\clearpage
\section{Applications}

In order to test the Grand Equilibrium method, it has been applied  
to the pure Lennard-Jones fluid, a binary, and a ternary Lennard-Jones
model mixture. The Lennard-Jones intermolecular potential acting between 
two particles of species $i$ and $j$ is defined by
\begin{equation}
u_{ij}(r) ~=~ 4\varepsilon_{ij} \left[\left(\frac{\sigma_{ij}}{r}
\right)^{12} -\left(\frac{\sigma_{ij}}{r}\right)^{6}\right], \label{ulj}
\end{equation} 
where $r$ is the intermolecular distance and $\sigma_{ij}$ and 
$\varepsilon_{ij}$ are the specific size and energy parameters.
In the case of a Lennard-Jones mixture, the full set of size and energy 
parameters for all interactions has to be specified. 
Unlike interaction parameters are often expressed on the basis of the like 
interaction parameters by
\begin{equation}
\sigma_{ij} ~=~ (\sigma_{i} + \sigma_{j})/2, 
\label{lbsig}
\end{equation}
and
\begin{equation}
\varepsilon_{ij} ~=~ \xi_{ij}\cdot\sqrt{\varepsilon_{i} \cdot \varepsilon_{j}},~ 
\label{lbeps}
\end{equation}
where $\xi_{ij}$ is a binary interaction parameter.

\bigskip
\bigskip
\subsection{Pure Lennard-Jones fluid}

The pure Lennard-Jones fluid, being a very simple but realistic molecular 
model, has been studied extensively by many authors. 
Its vapour-liquid equilibria are known with high accuracy 
\cite{lotfi,letter,okumura}, so that it is a favourable test fluid here. 

In a pure fluid the size and energy parameters are constant for all
interactions, so that $\sigma_{ij}=\sigma$ and $\varepsilon_{ij}=\varepsilon$.
All thermodynamic quantities are reduced in the usual way: 
temperature $T^*=Tk/\varepsilon$, pressure $p^*=p\sigma^3/\varepsilon$, 
density $\rho^*=\rho\sigma^3$, configurational energy $u^*=u/\varepsilon$,
configurational enthalpy $h^*=h/\varepsilon$, isothermal compressibility 
$\beta_T^*=\beta_T\varepsilon/\sigma^3$,
and residual chemical potential as given by equation
(\ref{mures}).

For a test of the Grand Equilibrium method on the pure Lennard-Jones 
fluid, no liquid simulations have been performed in the present work. 
As for the liquid phase simulations, only standard methods are needed, i.e. 
NpT ensemble and Widom's test particle insertion, so results from the 
literature \cite{lotfi} which are known to be accurate were used:
Table 1 comprises this data. 

Vapour simulations following the Grand Equilibrium method have been performed
based on these results. The technical details can be found in the appendix.
The vapour-liquid equilibrium results obtained with the Grand Equilibrium 
method are given in table 2. That table also contains
the widely accepted results of Lotfi {\it et al.} \cite{lotfi}, which were 
obtained using the NpT plus test particle method \cite{fischer}. 
The statistical uncertainties of the present vapour-liquid equilibrium results on the dew line were
determined by block averaging \cite{fincham8645} of the vapour simulations.
In this way, the influence of the uncertainties of the chemical potentials 
and the partial molar volumes from the liquid simulation is neglected. 
It should be emphasized that this leads to an underestimation.
For a repeatability study, see section 3.2. 
The statistical uncertainties of the vapour-liquid equilibrium results 
on the bubble line were determined by block averaging \cite{fincham8645} of 
the liquid simulation and the utilisation of the error propagation law.

It can be seen from table 2 that all data points for vapour 
pressure, bubble density, and dew density from the present work and those
obtained by Lotfi {\it et al.} \cite{lotfi} agree within their very small 
statistical uncertainties. The uncertainties of the Grand 
Equilibrium method are lower throughout than those of 
the NpT plus test particle method. In our opinion, this is due to the above
mentioned neglecting of the uncertainties of the chemical potentials 
and the partial molar volumes from the liquid simulation, since
this is an important source of error and the quality of the results of both 
methods should be similar.

\bigskip
\bigskip
\subsection{Binary Lennard-Jones mixture}

The Grand Equilibrium method was also tested on a binary
Lennard-Jones model mixture, of which the vapour-liquid equilibrium is 
well known. The parameters of the investigated symmetrical 
binary model mixture A+B are $\sigma_A=\sigma_B$, 
$\varepsilon_A=\varepsilon_B$, and $\xi_{AB}=0.75$, cf. equations 
(\ref{ulj})--(\ref{lbeps}). The thermodynamic
quantities are reduced in the same way as for the pure Lennard-Jones fluid,
using the parameters of the A--A interaction, $\sigma_A$ and 
$\varepsilon_A$. For comparison to the data obtained in the present work, 
results of the following authors are used:
Panagiotopoulos {\it et al.} \cite{pana2} and Tsangaris {\it et al.} 
\cite{tsangaris} who applied the Gibbs Ensemble Monte Carlo 
method (GEMC) as well as Boda {\it et al.} \cite{boda} who applied the Grand 
Canonical Monte Carlo method (GCMC). 

Binary vapour-liquid equilibrium data is calculated for the two temperatures 
$T^*=1$ and $T^*=1.15$. Table 3 comprises the simulation results for the 
liquid phase. It should be pointed out that all liquid simulations on one 
isotherm have been performed at the same arbitrarily chosen pressure 
$p^{l*}$. On the basis of these results, the vapour-liquid equilibria were 
determined by pseudo-$\mu$VT simulations and are given in table 4. The 
technical details of the simulations can be found in the appendix. 
The statistical uncertainties indicated in table 4 were estimated with the 
same procedure as described above for the pure fluid.

In order to investigate the statistical uncertainties of the Grand
Equilibrium method appropriately, the vapour-liquid equilibrium for
$T^*=1.15$ and $x_A=0.05$ have been calculated independently four times, 
cf. table 5. A better estimation of the uncertainties of the vapour-liquid 
equilibrium can be made by comparing these results. The uncertainties are 
for vapour composition $\delta y_A=0.002$, vapour pressure
$\delta p_{\sigma}^*=0.001$, bubble density $\delta\rho'^*=0.005$, dew 
density $\delta\rho''^*=0.002$, bubble enthalpy $\delta h'^*=0.03$, and 
dew enthalpy $\delta h''^*=0.02$. These values are higher throughout than
those obtained by the statistical analysis of the vapour runs,
which is easily explained by neglecting of the 
uncertainties of the chemical potentials and the partial molar volumes 
from the liquid simulation on the vapour-liquid equilibrium. 
They are still considerably lower than those reported by Panagiotopoulos
{\it et al.} \cite{pana2} for their method.

Different comparisons between the results of the present work obtained using 
the Grand Equilibrium method and GEMC 
results of Panagiotopoulos {\it et al.} \cite{pana2} and Tsangaris 
{\it et al.} \cite{tsangaris} as well as GCMC results by Boda {\it et al.}
\cite{boda} are given in figures 2--4. From the pressure--composition 
vapour-liquid equilibrium diagram in figure 2, it can be seen that the 
present data shows very little scatter, in the range of less than 0.01 in 
composition, and that it corresponds very well to the vapour-liquid 
equilibrium of the pure fluid B at $x_A=y_A=0$.
The agreement with the GEMC results \cite{pana2,tsangaris} is excellent, 
as almost all vapour-liquid equilibrium points agree with the 
GEMC results \cite{pana2,tsangaris} within the statistical uncertainties. 
The agreement with the GCMC results \cite{boda} at the lower temperature is 
good, especially on the dew line. But at the higher temperature considerable 
deviations occur. For a given pressure, the GCMC \cite{boda} compositions on 
both the bubble and dew line seem to be too low by 0.02--0.03. 

In figure 3 a pressure--density vapour-liquid equilibrium diagram is given. 
Again, the present data shows very little scatter and is consistent to the 
results for pure fluid B, for which the saturated densities can be found as 
the lowest point of each isothermal branch. The agreement to the 
GEMC results \cite{pana2,tsangaris} is excellent, as all 
points are within the statistical uncertainties of the GEMC 
results \cite{pana2,tsangaris}. The dew densities of the GCMC method \cite{boda} 
agree very well with the other data, but the bubble densities of \cite{boda}
especially at the higher temperature are somewhat too high. 

Figure 4 shows a pressure--energy density vapour-liquid equilibrium diagram, where the energy density
is defined by the product of the configurational energy and the
density. Due to the lack of appropriate literature data for $T^*=1.15$, only 
the isotherm $T^*=1$ is shown. Once more, it is found that the energy 
density obtained with the Grand Equilibrium method shows very little scatter 
and corresponds very well to the limiting 
pure fluid B, which is included as the lowest point of the present data. The
agreement with GEMC \cite{pana2,tsangaris} is again excellent. Consistent with 
the densities, the GCMC method \cite{boda} agrees well on the dew line, 
whereas on the bubble line, particularly for higher pressures, deviations 
are observed.

\bigskip
\bigskip
\subsection{Ternary Lennard-Jones mixture}

For a further test, the Grand Equilibrium method was applied 
to a ternary Lennard-Jones model mixture A+B+C, for which vapour-liquid equilibrium results from
Tsang {\it et al.} \cite{tsang} using the GEMC method, were available.

That ternary mixture is defined on the basis of equations 
(\ref{ulj})--(\ref{lbeps}) by $\sigma_A=\sigma_B=\sigma_C$,
$\varepsilon_B=0.75\varepsilon_A$, 
$\varepsilon_C=0.15\varepsilon_A$, and $\xi_{AB}=\xi_{AC}=\xi_{BC}=1$.
All thermodynamic quantities are reduced in the usual way, using the 
parameters of the A--A interaction, $\sigma_A$ and $\varepsilon_A$.

In the first step, liquid simulations for different compositions have been
performed at $T^*=1$ and $p^*=0.2$. The full set of these results is
given in table 6. The vapour-liquid equilibrium data is calculated by 
pseudo-$\mu$VT simulations in the vapour phase as described above, 
cf. table 7. The technical details of the simulations
are given in the appendix. The lower bounds of the statistical uncertainties 
of the vapour-liquid equilibrium, as given in table 7, were estimated by the same procedure as
described above for the pure fluid. It should be mentioned, that for each 
vapour-liquid equilibrium point only one simulation in the liquid phase and one simulation in the 
vapour phase is neccessary, regardless of the number of involved components. 

Figure 5 shows a ternary plot of the vapour-liquid equilibrium at $T^*=1$ 
and $p^*=0.2$. The results from the present work are
compared to the GEMC results of Tsang {\it et al.} \cite{tsang}.
It can be seen that both data sets are in excellent agreement.

\clearpage
\section{Conclusion} 

A new Grand Equilibrium method for the calculation of vapour-liquid 
equilibria is presented. 
It is aimed at mixtures, but can be applied to pure fluids straightforwardly.
The independent variables are temperature and liquid composition which is
favourable for bubble point calculations.

Two subsequent independent simulation runs have to be performed, first in
the liquid, then in the vapour, where standard simulation techniques 
are utilized. The method requires no prior knowledge of the vapour-lquid 
equilibrium, is easy to use, and performs reliably.

The Grand Equilibrium method has been applied successfully to the pure 
Lennard-Jones fluid, a binary, and a ternary mixture. 
Comparisons to results from other methods show that it yields 
reliable results with very little scatter.

\clearpage
{\bf Appendix}

For the liquid phase, NpT molecular dynamics simulations were performed, 
using the algorithm of Andersen \cite{andersen} and applying periodic 
boundary conditions as well as the minimum image convention. 
A cut-off radius of 4$\sigma_A$ was introduced, considering the long range
corrections. The equations of motion for 500 particles were 
solved with a fifth-order predictor-corrector method  using a time step of 
0.005 in usual units. Starting from a f.c.c. lattice,
the equilibration period was 5 000 time steps, whereof the first 1 000 
were carried out in the NVT ensemble. The length of the production run was 
100 000 time steps.
In order to calculate the chemical potentials and the partial molar volumes, 
Widom's test particle method \cite{widom6328} was utilised, by inserting 
1 000 test particles after each time step. The statistical uncertainties of 
the simulation results were determined according to the method of Fincham 
\cite{fincham8645}. 

For the vapour phase, pseudo-$\mu$VT Monte Carlo simulations were performed,
applying periodic boundary conditions and the minimum image convention.
The cut-off radius was again 4$\sigma_A$ and the long range
corrections were considered. The maximum displacement was set to 5\% of the 
simulation box length, which was chosen to yield on average 300 to 400
particles in the volume. After 1 000 initial NVT loops starting 
from a f.c.c. lattice, 9~000 equilibration 
loops in the pseudo-$\mu$VT ensemble were performed. One loop is 
defined here to be a number of attempts to displace particles equal
to the actual number of particles plus three insertion and three deletion
attempts. The length of the production run was 100 000 loops.

\clearpage
{\bf Acknowledgements}

\bigskip
The authors thank Mr. J\"urgen Stoll for fruitful discussions. 
We gratefully acknowledge financial support by Deutsche Forschungsgemeinschaft, 
Sonderforschungsbereich 412, University of Stuttgart.

% Bibliography

\clearpage

% Tables
\begin{table}[t]
\noindent
\caption{Thermodynamic properties in the liquid
phase of the pure Lennard-Jones fluid taken from Lotfi {\it et al.}
\cite{lotfi}.
The numbers in paranthesis denote the uncertainty in the
last digit.}
\label{1}
\bigskip
\begin{center}
\begin{tabular}{cc|ccc}\hline
$T^*$ & $p^*$ & $\tilde\mu$ & $\rho^*$ & $\beta_T^*$ \\ \hline
0.75 & 0.00 & -5.69\phantom{0} (3) & 0.8214           (4) & 0.091           (3) \\
1.00 & 0.03 & -3.823           (7) & 0.7018           (4) & 0.28\phantom{0} (2) \\
1.15 & 0.06 & -3.200           (4) & 0.6056           (6) & 0.6\phantom{00} (1) \\
1.25 & 0.10 & -2.882           (4) & 0.516\phantom{0} (2) & 2.3\phantom{00} (5) \\ \hline
\end{tabular}
\end{center}
\end{table}

\clearpage
   
\begin{table}[t]
\noindent
\caption{Comparison of vapour-liquid equilibria results of the pure 
Lennard-Jones fluid from the present work (Grand Equilibrium method)
and from Lotfi {\it et al.} \cite{lotfi} (NpT plus test particle method).
The numbers in paranthesis denote the uncertainty in the last digit.}
\label{2}
\bigskip
\begin{center}
\begin{tabular}{c|cccl}\hline
$T^*$ & $p_{\sigma}^*$ & $\rho'^*$ & $\rho''^*$ & \\ \hline
0.75 & 0.00261           (1) & 0.8216           (4) & 0.00359           (1) & present work\\
     & 0.00264           (7) & 0.8216           (4) & 0.0036\phantom{0} (1) & \cite{lotfi} \\
1.00 & 0.02500           (4) & 0.7008           (4) & 0.0296\phantom{0} (1) & present work\\
     & 0.0250\phantom{0} (2) & 0.7008           (4) & 0.0296\phantom{0} (3) & \cite{lotfi} \\
1.15 & 0.0597\phantom{0} (2) & 0.6054           (6) & 0.0728\phantom{0} (4) & present work\\
     & 0.0597\phantom{0} (4) & 0.6055           (7) & 0.0727\phantom{0} (8) & \cite{lotfi} \\
1.25 & 0.0971\phantom{0} (5) & 0.512\phantom{0} (2) & 0.136\phantom{00} (2) & present work\\ 
     & 0.097\phantom{00} (1) & 0.513\phantom{0} (3) & 0.134\phantom{00} (7) & \cite{lotfi} \\ \hline
\end{tabular}
\end{center}
\end{table}

\clearpage

\begin{table}[t]
\noindent
\caption{Thermodynamic properties in the liquid
phase of the binary Lennard-Jones mixture 
$\sigma_A=\sigma_B$, $\varepsilon_A=\varepsilon_B$, 
$\xi_{AB}=0.75$.
The numbers in paranthesis denote the uncertainty in the
last digit.}
\label{3}
\bigskip
\begin{center}
\begin{tabular}{c|cccccccr}\hline
$x_A$ & $\tilde\mu_A$ & $\tilde\mu_B$ & $v^*_A$ & $v^*_B$ & $\rho^*$ & $h^*$ & $\beta_T^*$ & $(\frac{\partial h^*}{\partial p^*})_{T,{\bf x}}$ \\ \hline
\multicolumn{9}{l}{$T^*=1$, $p^{l*}=0.04$}                              \\ \hline 
0.05 & -4.73\phantom{0} (1) & -3.87\phantom{0} (1) & 2.0 (1) & 1.3 (1) & 0.6892 (6)           & -5.637 (7)           & 0.29           (1) &  0.13           (6) \\
0.10 & -4.302           (8) & -3.908           (8) & 1.8 (1) & 1.2 (1) & 0.6744 (5)           & -5.428 (6)           & 0.31           (2) &  0.14           (7) \\
0.15 & -4.140           (8) & -3.931           (8) & 1.9 (1) & 1.3 (1) & 0.6618 (6)           & -5.262 (6)           & 0.37           (2) & -0.02           (7) \\
0.20 & -4.042           (8) & -3.940           (7) & 2.1 (2) & 1.5 (1) & 0.6506 (7)           & -5.11\phantom{0} (1) & 0.47           (2) & -0.3\phantom{0} (1) \\ \hline
\multicolumn{9}{l}{$T^*=1.15$, $p^{l*}=0.08$}                             \\ \hline 
0.05 & -4.641           (8) & -3.226           (6) & 3.0 (2) & 1.6 (1) & 0.587\phantom{0} (1) & -4.96\phantom{0} (3) & 0.84           (4) & -1.5\phantom{0} (2) \\ 
0.10 & -4.181           (5) & -3.270           (4) & 2.7 (2) & 1.3 (1) & 0.559\phantom{0} (1) & -4.68\phantom{0} (3) & 1.03           (6) & -1.9\phantom{0} (2) \\ 
0.15 & -3.952           (6) & -3.294           (4) & 3.8 (3) & 1.6 (1) & 0.534\phantom{0} (2) & -4.45\phantom{0} (5) & 2.0\phantom{0} (2) & -4.6\phantom{0} (5) \\ 
0.20 & -3.833           (6) & -3.306           (3) & 4.4 (4) & 1.3 (2) & 0.497\phantom{0} (3) & -4.17\phantom{0} (7) & 3.9\phantom{0} (4) & -9\phantom{.00} (1) \\ \hline
\end{tabular}
\end{center}
\end{table}

\clearpage

\begin{table}[t]
\noindent
\caption{Vapour-liquid equilibria of the binary Lennard-Jones mixture 
$\sigma_A=\sigma_B$, $\varepsilon_A=\varepsilon_B$, 
$\xi_{AB}=0.75$.
The numbers in paranthesis denote the uncertainty in the
last digit.}
\label{4}
\bigskip
\begin{center}
\begin{tabular}{c|cccccc}\hline
$x_A$ & $y_A$ & $p_{\sigma}^*$ & $\rho'^*$ & $\rho''^*$ & $h'^*$ & $h''^*$ \\ \hline
\multicolumn{7}{l}{$T^*=1$}  \\ \hline 
0.05 & 0.275 (2) & 0.0346 (1) & 0.6881           (6) & 0.0424           (1) & -5.638           (7) & -0.524           (2) \\
0.10 & 0.390 (2) & 0.0403 (1) & 0.6745           (5) & 0.0513           (2) & -5.428           (6) & -0.611           (2) \\
0.15 & 0.438 (2) & 0.0436 (1) & 0.6627           (6) & 0.0568           (2) & -5.262           (6) & -0.669           (3) \\
0.20 & 0.472 (2) & 0.0467 (2) & 0.6526           (8) & 0.0627           (3) & -5.11\phantom{0} (1) & -0.735           (4) \\ \hline
\multicolumn{7}{l}{$T^*=1.15$} \\ \hline 
0.05 & 0.158 (1) & 0.0722 (2) & 0.584\phantom{0} (1) & 0.0911           (4) & -4.95\phantom{0} (3) & -1.064           (5) \\
0.10 & 0.248 (2) & 0.0806 (3) & 0.559\phantom{0} (1) & 0.1051           (6) & -4.68\phantom{0} (3) & -1.158           (7) \\
0.15 & 0.308 (2) & 0.0890 (3) & 0.543\phantom{0} (2) & 0.124\phantom{0} (1) & -4.49\phantom{0} (5) & -1.33\phantom{0} (1) \\
0.20 & 0.341 (2) & 0.0958 (4) & 0.528\phantom{0} (5) & 0.144\phantom{0} (1) & -4.32\phantom{0} (7) & -1.50\phantom{0} (1) \\ \hline
\end{tabular}
\end{center}
\end{table}

\clearpage

\begin{table}[t]
\noindent
\caption{Vapour-liquid equilibria of the binary Lennard-Jones mixture 
$\sigma_A=\sigma_B$, $\varepsilon_A=\varepsilon_B$, 
$\xi_{AB}=0.75$ for the state point $T^*=1.15$ and $x_A=0.05$
from independent simulation runs.
The numbers in paranthesis denote the uncertainty in the
last digit.}
\label{5}
\bigskip
\begin{center}
\begin{tabular}{cccccc}\hline
$y_A$ & $p_{\sigma}^*$ & $\rho'^*$ & $\rho''^*$ & $h'^*$ & $h''^*$   \\ \hline
0.160 (1) & 0.0732 (3) & 0.588 (1) & 0.0931 (5) & -4.97 (3) & -1.081 (7) \\
0.159 (1) & 0.0727 (2) & 0.584 (1) & 0.0921 (5) & -4.96 (3) & -1.071 (7) \\
0.159 (2) & 0.0720 (3) & 0.583 (1) & 0.0910 (6) & -4.94 (3) & -1.060 (8) \\
0.158 (1) & 0.0722 (2) & 0.584 (1) & 0.0911 (4) & -4.95 (3) & -1.064 (5) \\ \hline
\end{tabular}
\end{center}
\end{table}

\clearpage

\begin{sidewaystable}
\centering
\caption{Thermodynamic properties in the liquid
phase of the ternary Lennard-Jones mixture 
$\sigma_A=\sigma_B=\sigma_C$, 
$\varepsilon_B=0.75\varepsilon_A$, $\varepsilon_C=0.15\varepsilon_A$, 
$\xi_{AB}=\xi_{AC}=\xi_{BC}=1$ at the temperature $T^*=1$ and
the pressure $p^{l*}$=0.2.
The numbers in paranthesis denote the uncertainty in the
last digit.}
\label{6}
\bigskip
%\begin{center}
\begin{tabular}{cl|cccccccccr}\hline
$x_A$ & ~~$x_B$ & $\tilde\mu_A$ & $\tilde\mu_B$ & $\tilde\mu_C$ & $v^*_A$ & $v^*_B$ & $v^*_C$ & $\rho^*$ & $h^*$ & $\beta_T^*$ & $(\frac{\partial h^*}{\partial p^*})_{T,{\bf x}}$ \\ \hline
0.972 & 0     & -3.61\phantom{0} (1) &            & -1.69\phantom{0} (1) & 1.2 (1) &         & 2.3 (1) & 0.7114 (4)           & -5.50 (2) & 0.202           (7) &  0.43           (4) \\
0.724 & 0.232 & -3.890           (8) & -3.869 (8) & -1.79\phantom{0} (1) & 1.4 (1) & 1.8 (1) & 3.1 (2) & 0.6663 (6)           & -4.79 (3) & 0.35\phantom{0} (2) &  0.05           (7) \\
0.432 & 0.492 & -4.282           (5) & -3.183 (5) & -1.954           (9) & 1.1 (1) & 1.8 (1) & 4.0 (2) & 0.5876 (8)           & -3.83 (6) & 0.74\phantom{0} (4) & -0.7\phantom{0} (1) \\
0.312 & 0.590 & -4.505           (3) & -3.007 (4) & -2.085           (8) & 0.6 (1) & 1.6 (1) & 5.1 (3) & 0.533\phantom{0} (7) & -3.33 (4) & 1.31\phantom{0} (7) & -1.8\phantom{0} (2) \\ \hline
\end{tabular}
%\end{center}
\end{sidewaystable}

\clearpage

\begin{table}[t]
\noindent
\caption{Vapour-liquid equilibria of the ternary Lennard-Jones mixture
$\sigma_A=\sigma_B=\sigma_C$,
$\varepsilon_B=0.75\varepsilon_A$, $\varepsilon_C=0.15\varepsilon_A$,
$\xi_{AB}=\xi_{AC}=\xi_{BC}=1$ at the temperature $T^*=1$.
The numbers in paranthesis denote the uncertainty in the
last digit.}
\label{7}
\bigskip
\begin{center}
\begin{tabular}{cl|ccccccc}\hline
$x_A$ & ~~$x_B$ & $y_A$ & $y_B$ & $p_{\sigma}^*$ & $\rho'^*$ & $\rho''^*$ & $h'^*$ & $h''^*$ \\ \hline
0.972 & 0     & 0.245 (3) & 0            & 0.1978 (8)           & 0.7111           (4) & 0.1893           (8) & -5.50 (2) & -0.297           (7) \\
0.724 & 0.232 & 0.205 (3) & 0.168    (2) & 0.207\phantom{0} (1) & 0.6680           (7) & 0.211\phantom{0} (1) & -4.79 (3) & -0.510           (8) \\
0.432 & 0.492 & 0.169 (4) & 0.369    (3) & 0.212\phantom{0} (1) & 0.593\phantom{0} (1) & 0.248\phantom{0} (2) & -3.84 (6) & -0.76\phantom{0} (1) \\
0.312 & 0.590 & 0.147 (5) & 0.457    (3) & 0.200\phantom{0} (1) & 0.532\phantom{0} (1) & 0.250\phantom{0} (2) & -3.33 (9) & -1.05\phantom{0} (2) \\ \hline
\end{tabular}
\end{center}
\end{table}

\clearpage

%\clearpage

% List of figures
\listoffigures
\clearpage

% Figures
\begin{figure}[ht]
\caption[Pressure, density, and vapour composition
over Monte-Carlo loops during the equilibration period.]{}

\label{X1}
\begin{center}
\includegraphics[width=116mm,height=153mm]{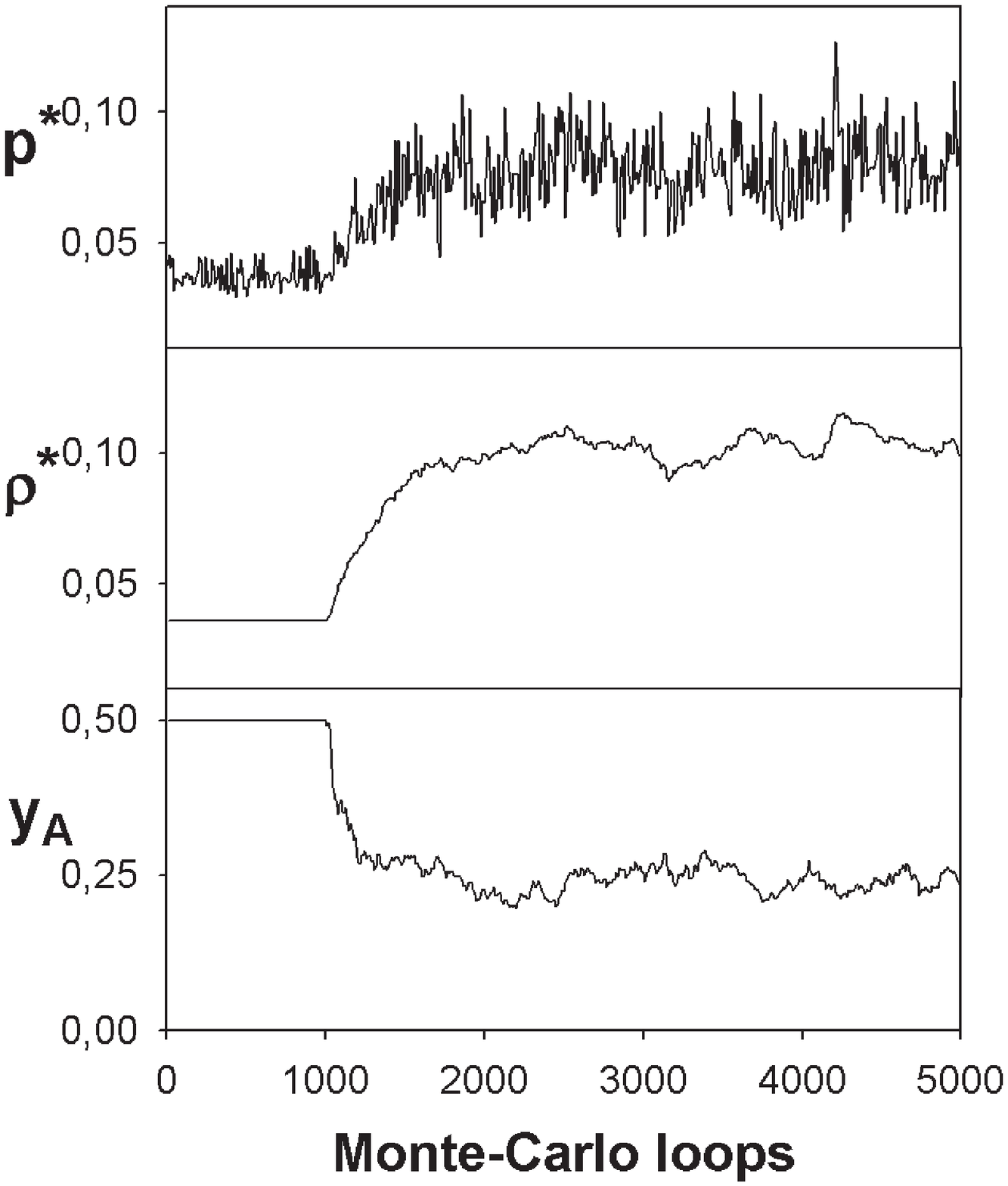}
\end{center}
\end{figure}

\begin{figure}[ht]
\caption[Vapour-liquid equilibrium diagram of the binary Lennard-Jones 
mixture $\sigma_A=\sigma_B$, $\varepsilon_A=\varepsilon_B$,
$\xi_{AB}=0.75$ at $T^*=1$ and $1.15$.
$\bullet$ Grand Equilibrium method (this work), 
$\blacktriangle$ GEMC results (Tsangaris {\it et al.} \cite{tsangaris}), 
$\blacktriangledown$ GEMC results (Panagiotopoulos {\it et al.} \cite{pana2}), 
--- GCMC results (Boda {\it et al.} \cite{boda}).]{}

\label{X2}
\begin{center}
\includegraphics[width=158mm,height=130mm]{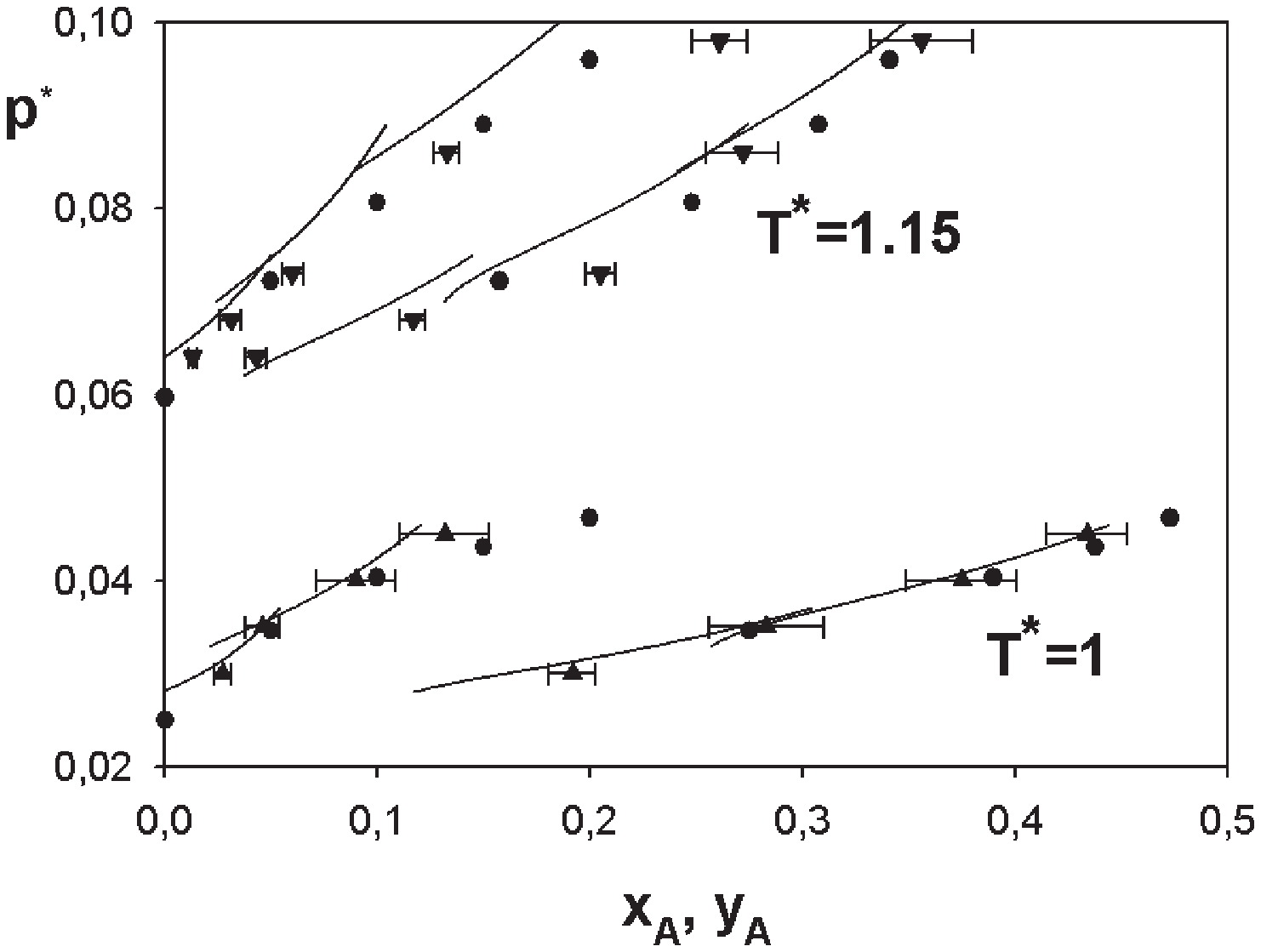}
\end{center}
\end{figure}

\begin{figure}[ht]
\caption[Vapour-liquid equilibrium diagram of the binary Lennard-Jones 
mixture $\sigma_A=\sigma_B$, $\varepsilon_A=\varepsilon_B$,
$\xi_{AB}=0.75$ at $T^*=1$ and $1.15$. Left: dew densities. 
Right: bubble densities. 
$\bullet$ Grand Equilibrium method (this work), 
$\blacktriangle$ GEMC results (Tsangaris {\it et al.} \cite{tsangaris}), 
$\blacktriangledown$ GEMC results (Panagiotopoulos {\it et al.} \cite{pana2}), 
--- GCMC results (Boda {\it et al.} \cite{boda}).]{}

\label{X3}
\begin{center}
\includegraphics[width=158mm,height=130mm]{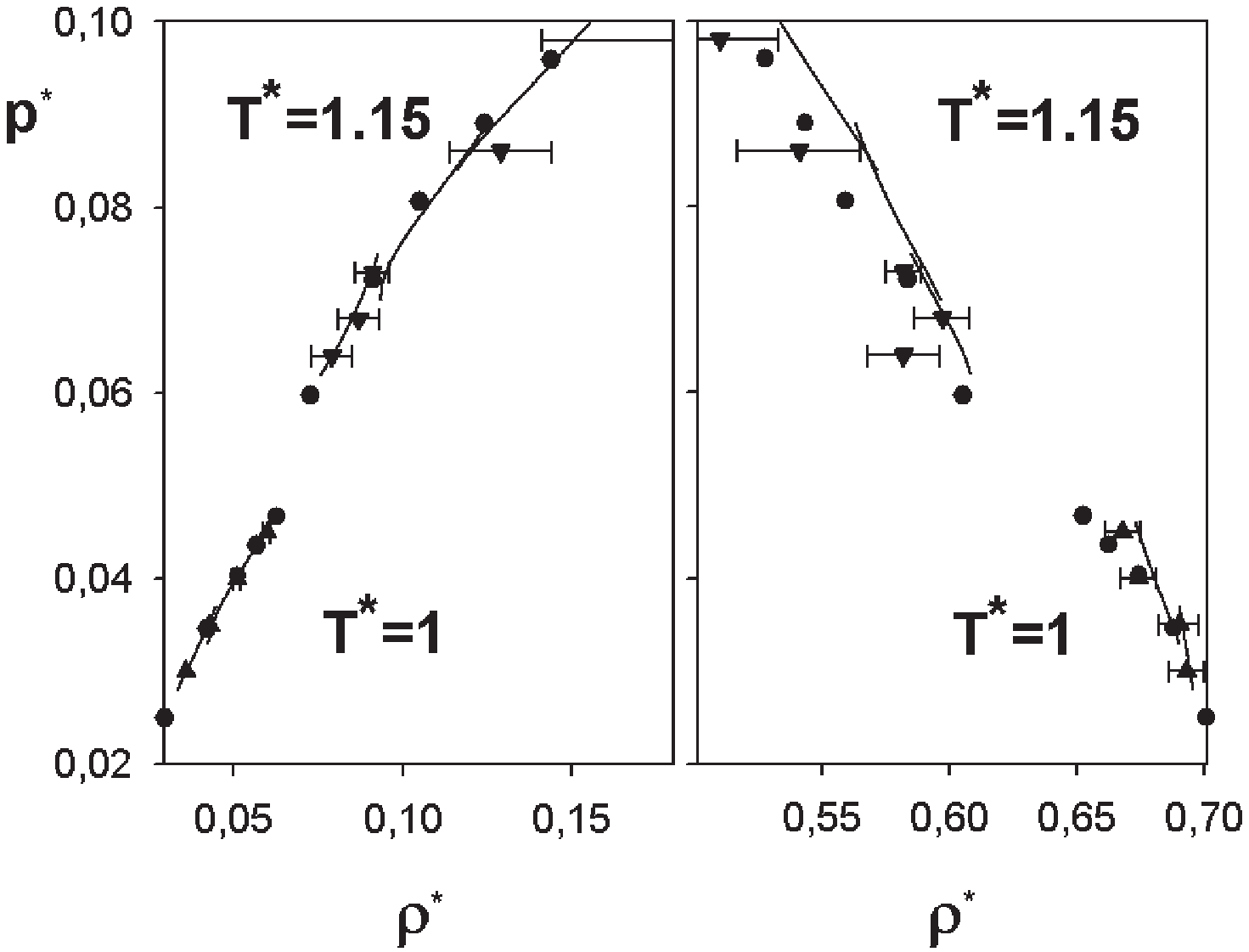}
\end{center}
\end{figure}

\begin{figure}[ht]
\caption[Vapour-liquid equilibrium diagram of the binary Lennard-Jones 
mixture $\sigma_A=\sigma_B$, $\varepsilon_A=\varepsilon_B$,
$\xi_{AB}=0.75$ at $T^*=1$.
Left: bubble energy densities. Right: dew energy densities.
$\bullet$ Grand Equilibrium method (this work), 
$\blacktriangle$ GEMC results (Tsangaris {\it et al.} \cite{tsangaris}), 
--- GCMC results (Boda {\it et al.} \cite{boda}).]{}

\label{X4}
\begin{center}
\includegraphics[width=158mm,height=130mm]{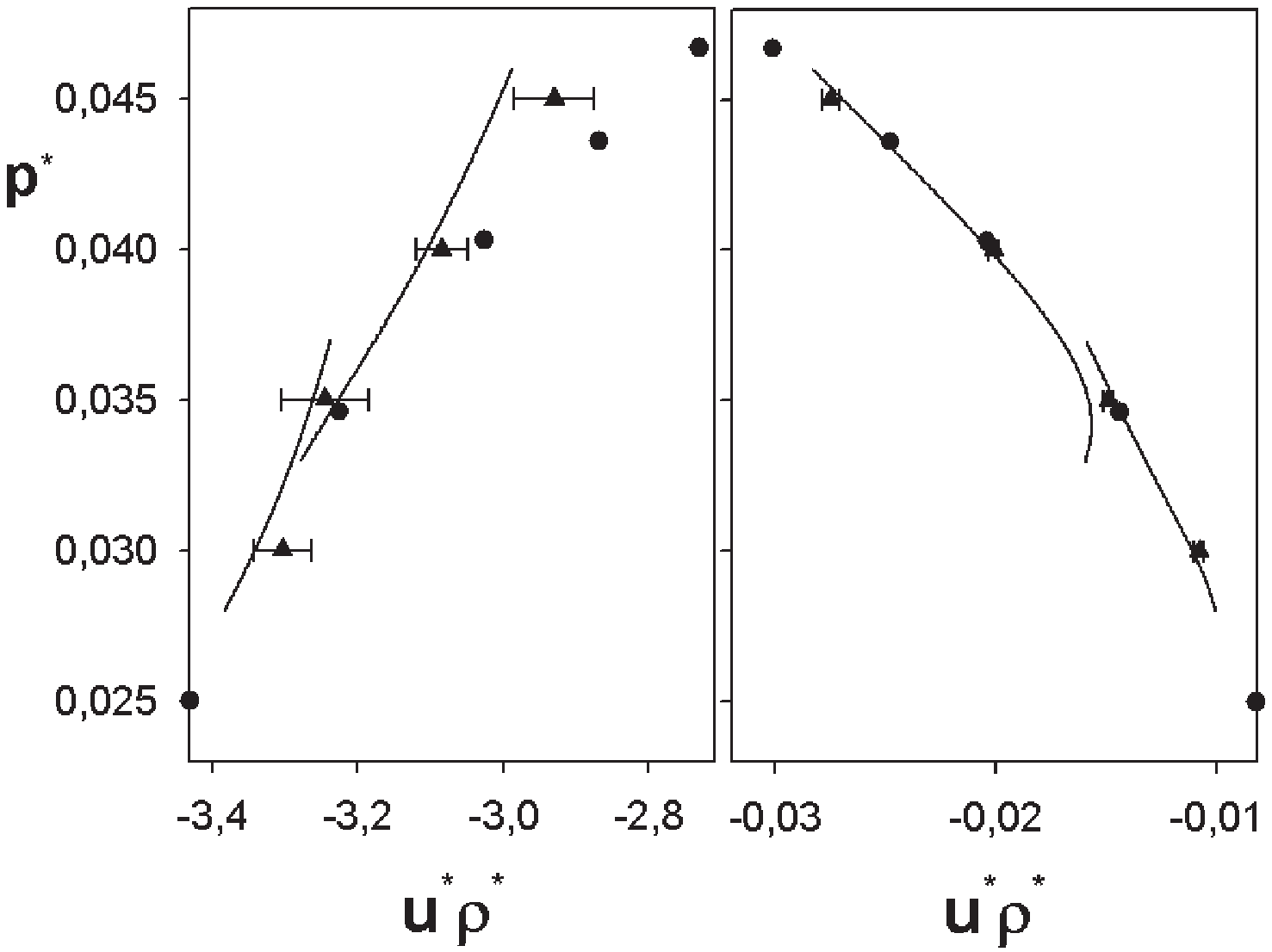}
\end{center}
\end{figure}

\begin{figure}[ht]
\caption[Vapour-liquid equilibrium of the ternary Lennard-Jones mixture 
$\sigma_A=\sigma_B=\sigma_C$, $\varepsilon_B=0.75\varepsilon_A$, 
$\varepsilon_C=0.15\varepsilon_A$, $\xi_{AB}=\xi_{AC}=\xi_{BC}=1$ at
$T^*=1$ and $p^*=0.2$.
$\bullet$ Grand Equilibrium method (this work), 
$\blacktriangle$ GEMC results (Tsang {\it et al.} \cite{tsang}),
--- conode.]{}

\bigskip
\bigskip
\bigskip
\label{X5}
\begin{center}
\includegraphics[width=158mm,height=145mm]{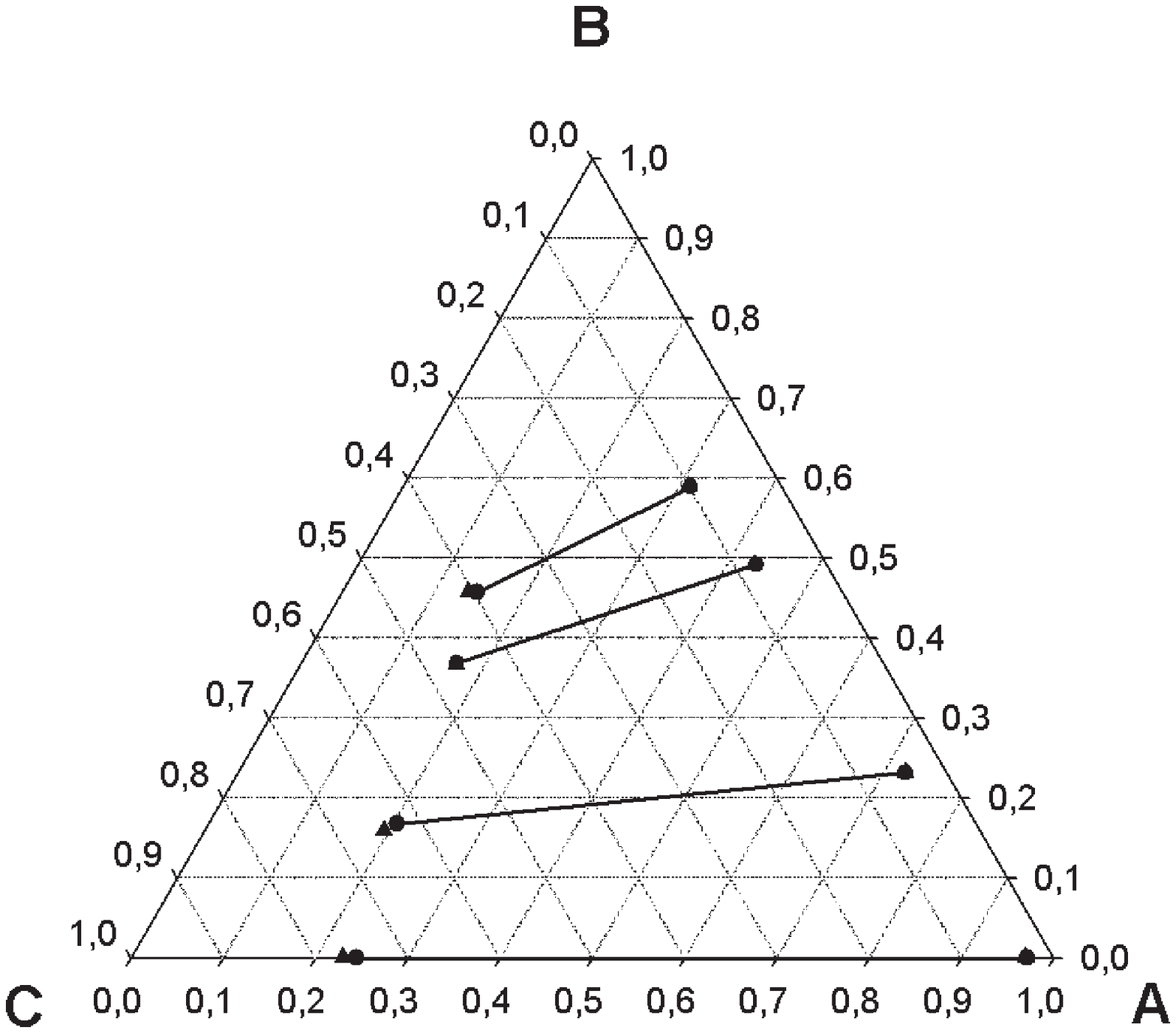}
\end{center}
\end{figure}

\clearpage


\begin{thebibliography}{99}
\bibitem{valleau1} Graham, I. S., and Valleau, J. P., 1990,  {\it J. Chem. Phys.}, {\bf 96}, 193.
\bibitem{valleau2} Valleau, J. P., 1991,  {\it J. Comput. Phys.}, {\bf 94}, 7894.
\bibitem{mcdonald1} McDonald, I. R., and Singer, K., 1967, {\it Discuss. Faraday Soc.}, {\bf 43}, 40.
\bibitem{potoff1} Potoff, J. J., and Panagiotopoulos, A. Z., 1998, {\it J. Chem. Phys.}, {\bf 109}, 10914.
\bibitem{kofke1} Kofke, D. A., 1993, {\it Molec. Phys.}, {\bf 78}, 1331.
\bibitem{kofke2} Kofke, D. A., 1993, {\it J. Chem. Phys.}, {\bf 98}, 4149.
\bibitem{boda0} Boda, D., Liszi, J., and Szalai, I., 1995, {\it Chem. Phys. Lett.}, {\bf 235}, 140.
\bibitem{boda1} Boda, D, Winkelmann, J., Liszi, J., and Szalai. I., 1996, {\it Molec. Phys.}, {\bf 87}, 601.
\bibitem{boda} Boda, D., Krist\'of, T., Liszi, J., and Szalai, I., 2001, {\it Molec. Phys.}, {\bf 99}, 2011.
\bibitem{fischer} M\"oller, D., and Fischer, J., 1990, {\it Molec. Phys.}, {\bf 69}, 463.
\bibitem{vrabec95} Vrabec, J., and Fischer, J., 1995, {\it Molec. Phys.}, {\bf 85}, 781. 
\bibitem{pana1} Panagiotopoulos, A. Z., 1987,  {\it Mol. Phys.}, {\bf 61}, 813.
\bibitem{pana2} Panagiotopoulos, A. Z., Quirke, N., Stapleton, M., and Tildesley, D. J., 1988, {\it Mol. Phys.}, {\bf 63}, 527.
\bibitem{pana3} Panagiotopoulos, A. Z., 1992,  {\it Fluid Phase Equilibria}, {\bf 76}, 97.
\bibitem{baus} Baus, M., Rull., L. F., and Ryckaert, J. P., 1995, {\it Observation, Prediction and Simulation of Phase Transitions in Complex Fluids}, (Dordrecht Kluver).
\bibitem{vrabec96} Vrabec, J., and Fischer, J., 1996, {\it AIChE J.}, {\bf 43}, 212. 
\bibitem{ungerer1} Ungerer, P., Boutin, A., and Fuchs, A. H., 1999, {\it Molec. Phys.}, {\bf 97}, 523.
\bibitem{ungerer2} Ungerer, P., Boutin, A., and Fuchs, A. H., 2001, {\it Molec.Phys.}, {\bf 99}, 1423.
\bibitem{escobedo1} Escobedo, F. A., 1998, {\it J. Chem. Phys.}, {\bf 108}, 8761.
\bibitem{escobedo2} Escobedo, F. A., 1999, {\it J. Chem. Phys.}, {\bf 110}, 11999.
\bibitem{hill} Hill, T. L., 1956, {\it Statistical Mechanics} (New York: McGraw-Hill Book Company).
\bibitem{widom6328} Widom, B., 1963, {\it J. Chem. Phys.}, {\bf 39},  2808.
\bibitem{heyes} Heyes, D. M., 1992, {\it Mol. Simul.}, {\bf 8}, 227.
\bibitem{voron1} Shevkunov, S. V., Martinovski, A. A., and Vorontsov-Velyaminov, P. N., 1988, {\it High Temp. Phys. (USSR)}, {\bf 26}, 246.
\bibitem{voron2} Lyubartsev, A. P., Martinovski, A. A., Shevkunov, S. V., and Vorontsov-Velyaminov, P. N., 1992, {\it J. Chem. Phys.}, {\bf 96}, 1776.
\bibitem{nezbeda9139} Nezbeda, I., and Kolafa, J., 1991, {\it Mol. Simul.}, {\bf 5}, 391.
\bibitem{allen} Allen, M. P., and Tildesley, D. J., 1987, {\it Computer simulations of liquids} (Oxford: Clarendon Press).
\bibitem{frenkel} Frenkel, D., and Smit, B., 1996, {\it Understanding Molecular Simulation} (San Diego: Academic Press).
\bibitem{sadus} Sadus, R. J., 1999, {\it Molecular Simulation of Fluid: Theory, Algorithms, and Object-orientation} (Amsterdam: Elsevier).
\bibitem{bodaneu} Kristof, T., Liszi, J., and Boda, D., 2002, {\it Molec. Phys.}, submitted.
\bibitem{stoll2001mix} Stoll, J., Vrabec, J., and Hasse, H., 2002, {\it AIChE J.}, submitted.
\bibitem{lotfi} Lotfi, A., Vrabec, J., and Fischer, J., 1992, {\it Molec. Phys.}, {\bf 76}, 1319.
\bibitem{letter} Vrabec, J., Lotfi, A., and Fischer, J., 1993, {\it Fluid Phase Equilibria}, {\bf 89}, 383.
\bibitem{okumura} Okumura, H., and Yonezawa, F., 2000, {\it J. Chem. Phys.} {\bf 113}, 9162.
\bibitem{fincham8645} Fincham, D., Quirke, N., and Tildesley, D. J., 1986, {\it J. Chem. Phys.}, {\bf 84}, 4535.
\bibitem{tsangaris} Tsangaris, D. M., and McMahon, P. D., 1992, {\it Mol. Simul.}, {\bf 9}, 223.
\bibitem{tsang} Tsang, P. C., White, O. N., Perigard, B. Y., Vega, L. F., and Panagiotopoulos, A. Z., 1995, {\it Fluid Phase Equilibria}, {\bf 107}, 31.
\bibitem{andersen} Andersen, H. C., 1980, {\it J Chem. Phys.}, {\bf 72}, 2384.
\end{thebibliography}
\end{document}